\documentclass[format=acmsmall, review=false, screen=true]{acmart}

\usepackage{booktabs} % For formal tables

\usepackage[ruled]{algorithm2e} % For algorithms
\usepackage{wrapfig}

\SetAlFnt{\small}
\SetAlCapFnt{\small}
\SetAlCapNameFnt{\small}
\SetAlCapHSkip{0pt}
\IncMargin{-\parindent}

\newcommand{\sys}{Crowd Coach\xspace}

\setcopyright{acmlicensed}
\acmJournal{PACMHCI}
\acmYear{2018} 
\acmVolume{2} 
\acmNumber{CSCW} 
\acmArticle{37} 
\acmMonth{11} 
\acmPrice{15.00}
\acmDOI{10.1145/3274306}

\received{April 2018} 
\received[revised]{July 2018}
\received[accepted]{September 2018} 

\begin{document}

\title[Crowd Coach: Peer Coaching for Crowd Workers' Skill Growth]{\sys:\\Peer Coaching for Crowd Workers' Skill Growth}

\author{Chun-Wei Chiang{$^1$}}
\affiliation{%
  \institution{HCI Lab, West Virginia University (WVU){$^1$}}
 % \streetaddress{HCI Lab}
 % \city{Morgantown}
 % \state{WV}
 % \postcode{26505}
  \country{USA}}
\email{cc0051@mix.wvu.edu}

\author{Anna Kasunic{$^2$}}
\affiliation{%
  \institution{Carnegie Mellon University (CMU){$^2$}}
%  \city{Pittsburgh}
  \country{USA}
}
\email{akasunic@andrew.cmu.edu}

\author{Saiph Savage{$^{1,2,3}$}}
\affiliation{%
 \institution{Universidad Nacional Autonoma de Mexico (UNAM){$^3$}}
 %Universidad Nacional Autonoma de Mexico (UNAM)$^{2}$
% \streetaddress{Rono-Hills}
 %\city{Morgantown}
 %\state{WV}
 %\postcode{26505}
 \country{Mexico}}
\email{saiph.savage@mail.wvu.edu }

\begin{abstract}
Traditional employment usually provides mechanisms for workers to improve their skills to access better opportunities. However, crowd work platforms like Amazon Mechanical Turk (AMT) generally do not support skill  development (i.e., becoming faster and better at work). While researchers have started to tackle this problem, most solutions are dependent on experts or requesters willing to help. However, requesters generally lack the necessary knowledge, and experts are rare and expensive. To further facilitate crowd workers' skill growth, we present Crowd Coach, a system that enables workers to receive peer coaching while on the job. We conduct a field experiment and real world deployment to study Crowd Coach in the wild. Hundreds of workers used Crowd Coach in a variety of tasks, including writing, doing surveys, and labeling images. We find that Crowd Coach enhances workers' speed without sacrificing their work quality, especially in audio transcription tasks. We posit that peer coaching systems hold potential for better supporting crowd workers' skill development while on the job. We finish with design implications from our research. 
\end{abstract}




\maketitle

\renewcommand{\shortauthors}{C.-W. Chiang et al.}

\section{Introduction}

In traditional employment, workers usually have the opportunity to grow their skills over time \cite{pew2013civic,williamson1998lifeworlds,duyff1999value, billett2001learning}. Work satisfaction theories have stressed the importance of fostering skill development in work environments to increase workers' motivations and contentment \cite{billett2001learning, ramlall2004review, ryan2000self, cartwright2006meaning}. This desire to develop one's skills is also present in crowd workers \cite{kaufmann2011more}. However, crowd markets in general have not been designed for skill growth \cite{bigham2017scopist,dontcheva2014combining,whiting2016crowd, suzuki2016atelier}. Consequently, crowd workers who wish to extend their skills must explore ways to train themselves outside crowdsourcing platforms \cite{kittur2013future}. However, given the low pay of crowd work \cite{paolacci2010running, berg2015income, durward2016crowd, thies2011paid, hara2018data}, requiring workers to use additional time and money for skill development is impractical \cite{kelliher2008better,van2017platform,rosenblat2016algorithmic,alkhatib2017examining}. To address this issue, scholars have recently started to explore models to enable skill growth while doing crowd work \cite{dontcheva2014combining, coetzee2015structuring, doroudi2016toward}. However, these models have relied heavily on requesters, whose time is limited, or domain experts, who are expensive \cite{suzuki2016atelier}. Requesters also have insufficient knowledge or motivation for helping workers \cite{doroudi2016toward,irani2013turkopticon,kulkarni2012turkomatic}. Consequently, these models are usually scarce and do not always address  workers' needs \cite{silberman2010ethics, Silberman:2010:SPH:1837885.1837891}. 
\begin{wrapfigure}{r}{7cm}
\vspace{0pc}
  \begin{center}
    \hspace{0pc}{\includegraphics[width=15pc]{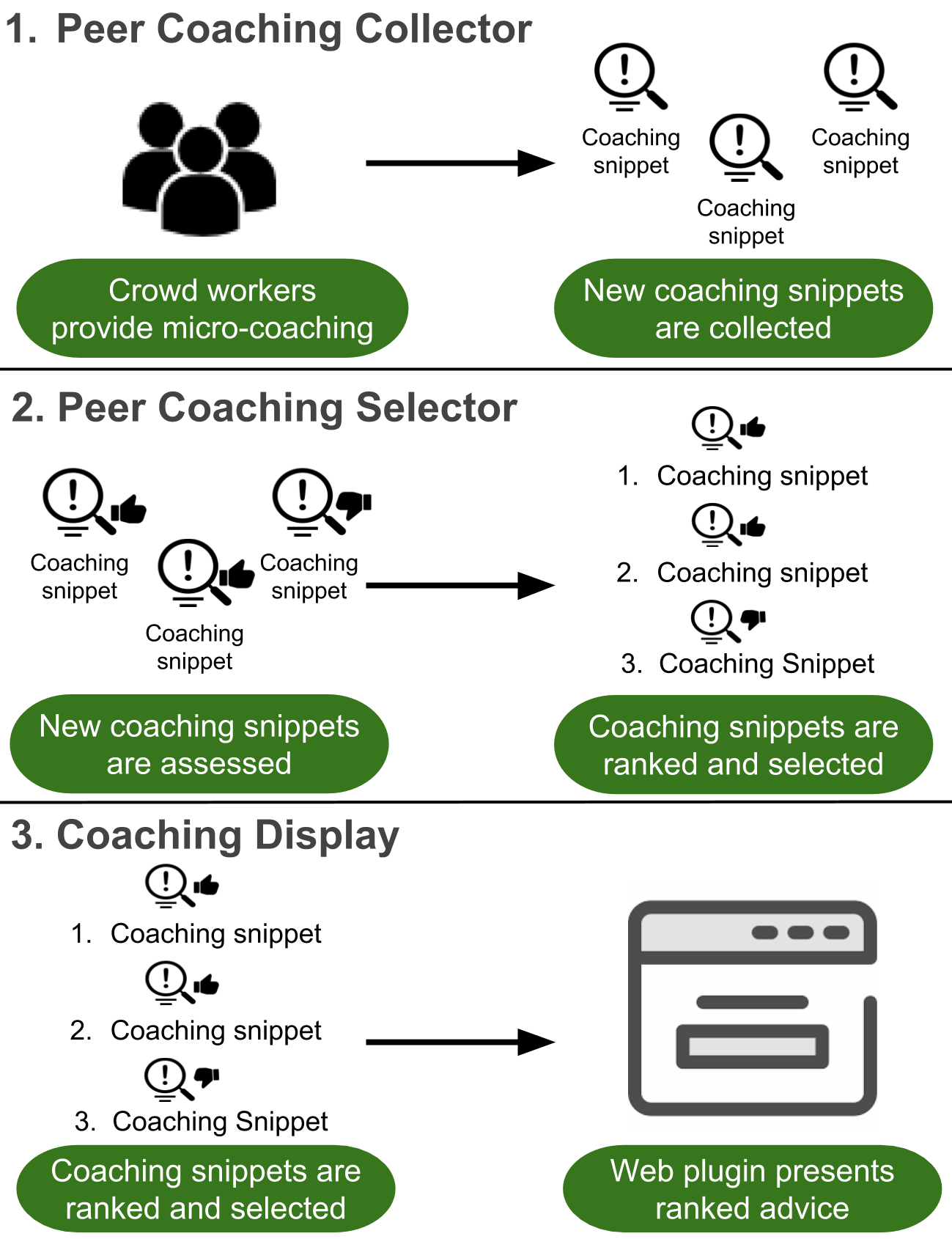}}
  \end{center}
\vspace{0pc}
  \caption{Overview of \sys's functionality.% consisting of: 1) Peer Coaching Collector; 2) Peer Coaching Selector; and 3) Coaching Display.
  }
\vspace{0pc}
\label{fig:workflow}
\end{wrapfigure} 

To enable skill development in crowd workers without requiring experts, we introduce the system \sys: a Chrome plugin that provides workers with short advice from peers while working on AMT. \sys provides two primary mechanisms to support skill development on crowd markets: (1) Micro-Coaching and (2) Selective Coaching.  Figure \ref{fig:workflow} presents an overview of our system. The Micro-Coaching interface encourages workers to provide short snippets of advice (coaching snippets) to each other while doing crowd work. This approach draws on ``Twitch crowdsourcing'' research \cite{vaish2014twitch} that focuses on enabling people to do micro-tasks rapidly and without them being a person's main activity. Based on these ideas, we specifically limit the coaching that workers provide to a length of 100 characters. This empowers workers to peer-coach while still doing their main job on AMT. In parallel, our Selective Coaching mechanism helps workers to follow the coaching that might best help them to develop their skills.  In specific, we focus on developing workers' ability to do specific tasks better over time  \cite{urciuoli2008skills}, which is typically measured in terms of the quality of work produced and the amount of time to complete the task \cite{angelo1993classroom,scopist}.

In this paper we contribute: 1) a system supporting crowd workers' skill development via peer coaching; 2) a mechanism to enable providing and receiving peer coaching while doing crowd work; 3) a field experiment demonstrating that \sys facilitates peer-coaching and benefits the crowd workers receiving the coaching; 4) a real world deployment showcasing how workers use \sys in the wild to better uncover the system's benefits and drawbacks. We believe \sys's method of using short, selective peer coaching has great potential for transforming the ways in which crowd workers can develop their skills while on the job. 

\section{Related Work}

\sys's system design is informed by four main research areas: 1) Skill development in crowd work; 2) Tools for improving crowdsourcing platforms; 3) Online Peer to Peer Support and Collaborations; 4) Peer learning.

%\subsection{Skill Development in Crowd Work}
{\bf 1. Skill Development in Crowd Work.} Both researchers and practitioners still struggle with enabling skill development on crowdsourcing platforms 
\cite{deng2013crowdsourcing, thies2011paid}. Enabling skill develop is important for several reasons. First, skill improvement enhances performance in crowd work \cite{kittur2013future}. Complex tasks also need skilled workers to complete them \cite{noronha2011platemate}. Having more specialized workers can facilitate the completion of more work and facilitate reaching more complex goals \cite{Chandler2013}. Second, skill development can help crowd workers to finish their tasks faster \cite{bigham2017scopist}. This can potentially aid crowd workers to obtain a higher hourly pay, as payment on AMT is typically based on workload (number of tasks completed) \cite{callison2014crowd, ikeda2016pay}. Third, ``learning new skills'' is one of the main motivators that crowd workers have for joining the crowd market \cite{kaufmann2011more}. Crowd workers expect that their experiences on crowdsourcing platforms will benefit their career development \cite{brabham2008crowdsourcing}. Nevertheless, most crowd workers lack the opportunity for improving or advancing their careers \cite{suzuki2016atelier}.     

To address this problem, researchers have developed tools to facilitate skill development in crowd work. Atelier for Upwork \cite{suzuki2016atelier} provided a mentorship model where experts helped crowd workers learn new skills; LevelUp for Photoshop \cite{dontcheva2014combining} integrated tutorials designed by experts to teach people designer skills while completing basic photograph improvement tasks. However, most of these tools require experts' time to help new workers, either via mentoring or by developing educational material such as tutorials. Yet, experts' time is limited and expensive \cite{suzuki2016atelier}. 

To reduce the dependency on experts, very recent research has started to explore tools that incorporate ideas from peer learning. Peer learning is a concept from cognitive psychology where students have the power to teach and learn from each other \cite{boud1999peer}. Prior work has uncovered that students who participate in peer learning perform better than students working on their own \cite{lundberg2003influence}, and were even better than students instructed by experts \cite{hinds2001bothered}. This learning method also appears to be suitable for crowd workers; researchers have found that crowd workers enhance the quality of their work when they reviewed and provided feedback to the labor of other crowd workers  \cite{coetzee2015structuring, zhu2014reviewing, doroudi2016toward}. However, these approaches assume that requesters will be willing to take the time to redesign their tasks to incorporate and facilitate the peer learning model. Under the current dynamics of AMT, it is likely that requesters lack: a) the knowledge on how to effectively design tasks and workflows to help workers \cite{kulkarni2012turkomatic}; b) the time and incentives to want to improve tasks and worker satisfaction \cite{irani2013turkopticon}. Moreover, these studies made a trade-off between improving workers' labor quality and helping workers to reduce their  completion time \cite{kulkarni2012turkomatic}. Prior research focused on helping workers produce higher quality work but with the cost of increasing their completion times. For instance, very recent investigations explored workflows where novice workers reviewed each other's work by writing long writeups about each other's labor (800 to 1,000 characters long) \cite{doroudi2016toward} or engaging in detailed discussions with each other \cite{zhu2014reviewing}. While such workflows did help crowd workers to improve their performance, some workers also had to invest 25\% more time in writing, reading and digesting each other's advice. Most crowd workers might not have the time to engage in such lengthy reviewing activity in addition to their main work. Morever, most workers are more interested in enhancing their skills to ultimately improve their salary than in increasing their work quality \cite{mason2010financial, martin2014being}. As explained above, increasing crowd workers' completion time per task can decrease workers' hourly wage. We believe that crowd workers might also be able to improve while on the job without sacrificing quality over completion time by providing quick, but selective peer advice. We consider this selective advice to be similar to coaching that is usually very direct and precise. Kittur et al. \cite{kittur2013future} envisioned that crowd markets could be a direct medium for skill development. \sys builds on this idea by helping workers to develop their skills without requiring experts to invest their time or requesters to redesign the tasks. Instead, \sys uses peer micro-advice to guide crowd workers to improve their performance, especially in working time without sacrificing quality. \sys also provides a quality assessment loop that dynamically finds high-quality peer advice (selective advice). 

%I can take a closer look at background later. For now, just pinning some sources that may be useful, depending on framing. 
%^also cite: sarasua2013microtask (about HITs based on CV)
%^may also be useful, but perhaps more related to other project: sarasua2014crowd (about "captures crowd workers’ interests, qualifications and work history, as well as requesters’ information")
%Also a number of sources about framing learning through work cited in \cite{billett2001learning} that might be useful
%Might want to frame in terms of motivation literature (to argue for why it's important to allow crowd workers to improve skills through their work)-- can \cite{kaufmann2011more} here, worker's motivation in crowdsourcing model. Can focus on idea of delayed payoffs and human capital advancement. Through this model, can also frame in terms of human capital advancement (so may want to cite other work related to human capital)
%also to consider: rogstadius2011assessment (crowdworker motivation improved when framed as helping others)

{\bf 2. Tools for Improving Crowdsourcing Platforms.} 
%\subsection{Tools for Improving Crowdsourcing Platforms.}
Due to the inequalities between requesters and workers in AMT, workers usually obtain unfair treatment \cite{irani2013turkopticon, irani2016stories}. Researchers and practitioners have been exploring different browser extensions to improve the working conditions of crowd workers and reduce the inequalities. The web plugin of Turkopticon allows crowd workers to evaluate requesters after finishing their tasks \cite{irani2013turkopticon}. Turkopticon has become a useful tool for reducing the information asymmetry gap between workers and requesters (on AMT only requesters could officially rate workers). Other tools have explored enabling workers to help each other find higher hourly paying jobs to offset the fact that most crowdsourcing platforms do not provide an estimated working time of each task \cite{callison2014crowd}. TurkBench proposed a market visualization tool to help crowd workers to better 
manage their tasks \cite{hanrahan2015turkbench}. These novel tools have improved the working environment of crowd workers. In a similar fashion, \sys aims to improve crowd workers' labor conditions by facilitating on-the-job skill development.

%{\bf 3. Eliciting Specialized Information from People Online.}
{\bf 3. Online Peer to Peer Support and Collaborations.}
Recently, we have seen the emergence of systems that coordinate peers of online strangers to share useful information with each other \cite{nichols2012asking}. Several human computation workflows have successfully driven strangers to share their knowledge to help others learn \cite{Weir:2015:LSL:2675133.2675219}. These studies have found that online strangers can indeed provide quality information \cite{nichols2013analyzing}, even when asked by bots \cite{savage2016botivist}. Related work has also investigated how receiving feedback or advice from peers affects the quality of online work produced in peer production communities \cite{He:2014:EIF:2556420.2556484,Dittus:2017:PPF:3171581.3134675}, such as Wikipedia \cite{Zhu:2013:EPF:2470654.2481311,Schneider:2014:ADP:2641580.2641614,Balestra:2017:IMP:3025453.3026057}.  
%Researchers have also started to investigate the type of feedback that is possible to manually obtain from different online sites, especially crowd markets, social networks, and forums \cite{yen2016social,kou2017supporting}. However, all of these settings assumed The problem however, is that in these platforms much time is spent interpreting what the learner produced \cite{kou2017supporting}.
We motivate the design of \sys on some of the key findings of this previous research: it is possible to drive online strangers to provide useful information for  others \cite{nichols2013analyzing,yen2016social}, and receiving peer advice can enhance a person's work \cite{Zhu:2013:EPF:2470654.2481311}. We hypothesize that if we make it simple enough to provide peer advice, we will be able to orchestrate the coaching of crowd workers for improving their skills while on the job. %integrate guidance we could orchestrate experts to effectively critique the creative work of learners they have never met before at scale.  

{\bf 4. Peer Learning.}
In the design of \sys we posit that we can coordinate crowd workers to coach others via peer advice. Prior work has showed that peer advice can help students to improve their grades \cite{falchikov2000student,kulkarni2013peer}. Usually, this is because peer advice can help students to: (1) focus on the aspects of their work that will lead to the highest grade increase \cite{kluger1996effects}; (2) better identify why they have poor performance \cite{balcazar1985critical}; or (3) encourage students to have higher work standards after seeing the great strides made by their peers \cite{hattie2007power,latham1991self}. More recent research has designed and studied peer learning systems that maximize these advantages. Kulkarni et al. \cite{kulkarni2015peerstudio} studied systems to provide rapid peer advice to MOOC learners. Yoon et al \cite{Yoon:2016:RDC:2818048.2819951} studied multi-modal peer advice systems. Yet most related work has studied peer learning within formal educational settings \cite{carlson2003calibrated,tinapple2013critviz,kulkarni2013peer}. We expand this research and study integrating peer advice in informal learning environments, specifically crowd markets. Our work also investigates the potential of enabling peer feedback while the individuals are doing  another main activity, specifically crowd work.

\section{\sys}
Motivated by the challenges of enabling crowd workers' skill development without depending on experts or requesters, our research translates peer learning models into the general design mechanism of ``crowd coaching.'' In crowd coaching, workers can provide and access peer advice (coaching snippets) to improve their skills. While there are many ways workers could give and obtain peer coaching, we focus on coaching that could take place while on the job. We considered it was important to minimize the time that crowd workers spend outside AMT to reduce the instances where workers are not receiving wages. We therefore envisioned a web plugin solution that would allow people to continue on AMT and avoid disrupting their normal work routines. We also considered that the coaching would not be the primary task that individuals are doing. Therefore, we design both the activities of coaching and receiving coaching to be lightweight and avoid distracting people. We integrate into our design ideas from ``Twitch crowdsourcing'' research \cite{vaish2014twitch} where people do micro-tasks without disrupting their main activity. For this purpose, we specifically frame \sys's design around: (i) Availability:  ability to evoke peer coaching with a click; (ii) Low Cognitive Load: allow crowd workers to coach and obtain coaching without the activity becoming a distraction to their main job. Given crowd workers' labor conditions \cite{hara2018data}, it was also important to integrate into our design: (iii) Paid Training: empower workers to personally improve their skills while earning money from their main job.

\begin{figure}
\centering
  \includegraphics[width=.9\columnwidth]{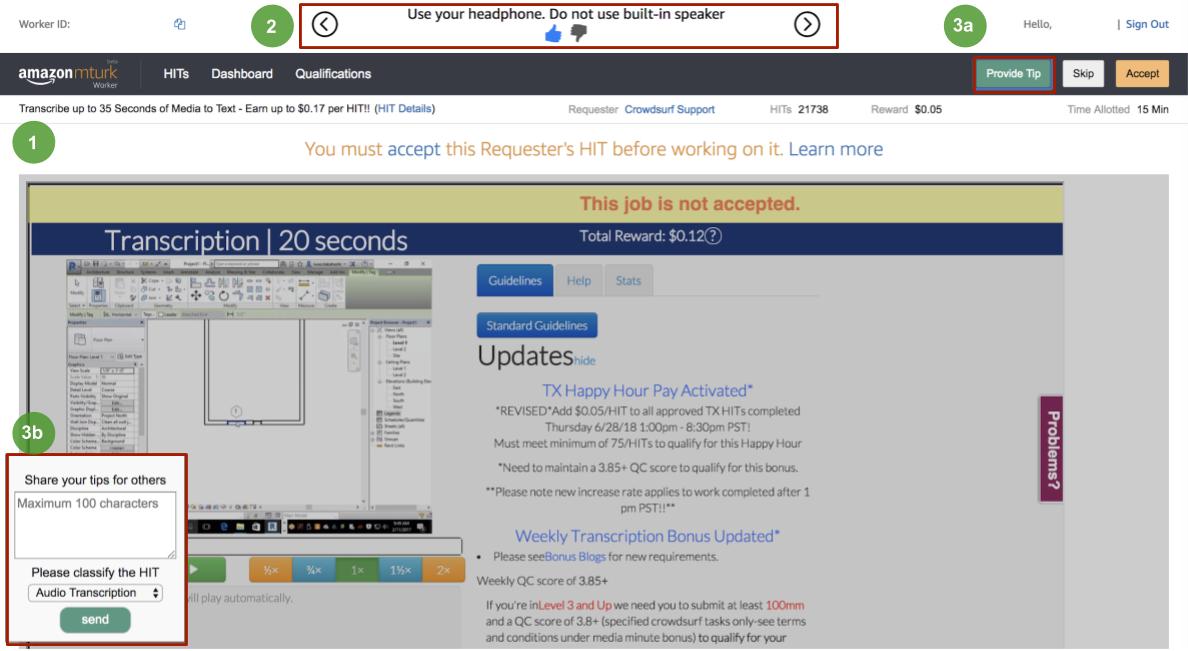}
  \caption{Screenshot of our system in action. \sys enables workers to develop their skills while on the job by: (1) integrating directly into AMT hits; (2) presenting selected coaching snippets to workers; and (3) allowing workers to easily coach. }
  \label{fig:interface}
\end{figure}

To enable these points, \sys has a: 1) Peer Coaching Collector, 2) Peer Coaching Selector (to select the coaching that is most useful), and 3) Coaching Display (to present useful coaching). Figure \ref{fig:workflow} presents an overview of \sys and Figure \ref{fig:interface} showcases an overview of our system's interface. 

{\bf 1. Peer Coaching Collector.} The Peer Coaching Collector enables workers to coach others by inputting advice on how to improve at particular HITs. The collector lives as a plugin on AMT where with a click, workers can become coaches to their peers and provide short coaching snippets on how to become faster and maintain good work quality for specific tasks (especially for the tasks that  workers, i.e., coaches, are currently doing within their main job). Unlike previous research \cite{doroudi2016toward} that leads workers to give lengthy information, \sys encourages micro-advice. We view this micro-advice as similar to what a coach would briefly yell to players during a training to help them personally improve. We thus call this advice ``coaching snippets''. For users of the plugin (Google Chrome extension), \sys shows a small ``provide tip'' button on the AMT status bar. Upon clicking the button, workers see a small pop-up window where they can provide their coaching snippets. Notice that this facilitates our design principle of  ``availability to coach.'' We also wanted to limit the cognitive load that coaching imposes on already occupied workers. We thus limit the length of people's coaching to 100 characters and try to make the coaching as simple as possible (we set the characters limit to 100 characters through trial and error). In the pop-up window, workers simply select the type of tasks for which their coaching snippets is relevant and then type their coaching snippet. This design enables us to match coaching to particular tasks without imposing a large amount of new labor on workers. Notice that although AMT tasks do provide description and titles, and we could use topic modeling or NLP techniques to categorize the tasks, prior work has shown that such written information is not always the most relevant to classify tasks \cite{hara2018data}. We therefore focused our efforts towards a crowdsourced solution. Based on prior work \cite{hara2018data, ross2009turkers}, \sys considers that 8 main types of tasks exist on AMT and asks workers to classify tasks into either: Audio transcription, Categorization, Data Collection, Image Transcription, Image Tagging / Labeling, Surveys, Writing and Other.  %\sys's Peer Coaching Collector asks crowd workers to classify tasks into one of these eight specific HIT types. %Through this process, we link micro-advice (coaching) to specific types of tasks. 

{\bf 2. Peer Coaching Selector.} For each type of AMT task, the Peer Coaching Collector returns a long queue of coaching snippets. However, some coaching snippets may not be that useful for workers' personal growth. Especially with a large number of snippets, relevant ``advice gems'' might get lost in the muck. To overcome this issue, we have a Peer Coaching Selective module that focuses on identifying the best coaching snippets for a task. 

The module first asks workers to micro-assess a coaching snippet via upvotes or downvotes.  
(see Figure \ref{fig:interface}). \sys tries for workers to micro-assess coaching related to tasks that workers are currently doing.  When a coaching snippet gets assessed, it will win or lose ``credits'' (depending on whether it was upvoted or downvoted). These credits are pro-rated based on the reputation of the worker assessing. The system calculates the reputation of workers based on: a) their experiences on AMT (number of tasks completed); and b) how similar their micro-assessment is to that of other workers. If a person's micro-assessment deviates too much from what others have input, we consider it as an ``alternative'' assessment. Using this metric we classify micro-assessments as either ``mainstream'' or ``alternative,'' and rank mainstream coaching based on peer ratings. 
%In \sys, workers can customize whether they receive  mainstream or non-mainstream coaching, and can view advice from workers with different experience levels. Through this mechanism, we enable workers to receive mainstream coaching (mainstream) or coaching with more unique perspectives (non-mainstream), as well as coaching from people with varying experience levels (whose coaching can be mainstream or non-mainstream). 
By default, \sys presents first highly ranked mainstream coaching snippets to workers.
% Figure \ref{fig:interface} presents an overview of how workers see this advice. 
Through its voting mechanism, \sys can thereby select coaching that peers found useful.  

{\bf 3. Coaching Display.} This component focuses on presenting coaching that will help workers to develop their personal skills while on the job. For a given task, the Coaching Display presents to workers four associated coaching snippets that were ranked highest on the list. If workers want to read more coaching snippets, they can click the left or right button to view more. To ensure that new coaching snippets have the opportunity to be evaluated, \sys always mixes new coaching snippets that needs micro-assessments into the list of high ranking coaching snippets. This dynamic helps to make sure that the new coaching snippets have the opportunity to be voted. Notice that workers can also choose to add their own coaching for a given task. \sys keeps track of the coaching that has already been shared with workers to avoid redundancy. Figure \ref{fig:interface} presents how the coaching is displayed to workers and how workers coach peers. 

\section{Evaluation}
This paper hypothesizes that receiving selected peer coaching can help workers get started in building their skills. Similar to \cite{doroudi2016toward}, we measure skill growth in terms of an increase in workers' speed and labor quality. To test this hypothesis and to understand the type of work that is well or poorly supported by \sys in the wild, we conducted: (1) a controlled field experiment;  and a (2) real world deployment of \sys. The controlled field experiment allows us to study how \sys might facilitate crowd workers' skill growth. The deployment allows us to investigate natural usages of \sys to probe the strengths and weaknesses of our system.

\subsection{Controlled Field Experiment}
The goal of our field experiment was to compare our peer coaching approach with other conditions to evaluate its effectiveness in developing crowd workers' own skills. We considered 3 conditions: 1) workers do tasks without receiving any type of peer coaching [control condition]; 2) workers do tasks while receiving random coaching snippets[random snippets condition]; 3) workers do tasks while receiving selected peer coaching [\sys condition]. 

Given that for our experiment we needed to measure participants' work quality, we focused on skill development for labor that was not open-ended and whose quality we could more easily measure. We specifically focus on audio transcription tasks whose quality is directly measured with people's transcription accuracy. Not only are audio transcription tasks one of the most common AMT tasks \cite{difallah2015dynamics}, but in addition, becoming good at audio transcription can substantially increase a person's wages.  Transcribers typically earn \$0.01-0.02 USD per sentence they transcribe \cite{novotney2010cheap}, which could potentially translate to high wages if a worker is fast (and accurate) enough. Specializing in audio transcription could allow crowd workers to command higher wages, as audio transcription is in high demand. Written records of court proceedings and captions for live television events, such as the news, sports, and political speeches all require real-time audio transcription. Audio transcription skills are thereby highly specialized, highly valued, and well paid, earning up to \$300 per hour outside AMT. Building audio transcription skills on AMT could thereby help crowd workers expand their horizons and increase their earnings. Given all of this, we considered it valuable to study \sys's effectiveness in improving workers' audio transcription skills, especially for the case of novice workers.  We considered that novices were the ones who could benefit the most from systems like \sys as it can be difficult to learn the ropes of the AMT ecosystem while also developing skills. Our field experiment thus focuses on investigating whether \sys helps novice crowd workers improve their audio transcription skills. 
%In our field experiment we thus focus on studying whether \sys can especially help workers to develop their audio transcription skills. 

\subsubsection{Field Experiment: Method}
We studied novices' completion time and work quality for 3 different audio transcription tasks under one of our 3 study conditions. We considered that novice workers were both workers who were new to AMT and inexperienced in audio transcription tasks.  %tasks with the help of \sys and compared their performance with novice workers completing the same labor without \sys. 
We first recruited crowd workers who had completed less than 100 HITs on AMT. Next, we identified which of these workers were also novices in audio transcription. For this purpose, we asked potential participants to do a pre-test (which consisted of completing real world audio transcription tasks). We included in our study only the workers who finished such tasks in a similar time and with similar work quality (accuracy). We recruited a total of 90 novice AMT workers, and randomly divided 30 participants into one of our 3 experimental conditions. Participants in each condition were assigned the same audio transcription tasks with the same order. We collected the coaching snippets and micro-assessments of the coaching snippets before the experiment. Workers for each task were shown the same random coaching snippets or the same selected coaching snippets (depending on their condition.) Workers were paid \$0.6 for completing an audio transcription tasks (\$1.8 in total). We paid workers \$0.6 when they accepted and worked on the first audio transcription task, and gave another \$0.6 as bonus when they completed one more task. Workers could dropout whenever they wanted and get paid for the work they had finished. Tasks were sourced from real world audio transcription HITs on AMT and had similar difficulty:  participants had to transcribe around 28 seconds of audio with similar levels of background noise, and with an average speaking rate of 165 word-per-second. To better trace participants' performance from task to task, we had participants interact with an AMT-like website we built. The website recorded workers' retention rate, completion time and accuracy for each task. To measure time to complete a task, we measured the time when a worker first accessed the task as the start time, and the time when workers submitted their labor (transcription) as their finish time. To study accuracy, we calculated the word error rate (WER) produced by each worker for each transcription, a commonly used metric to assess performance in audio transcription \cite{bigham2017deaf}. Participants also completed a survey about their experiences in each condition.

\subsubsection{Field Experiment: Result}

\begin{figure}
\centering
  \includegraphics[width=0.93\columnwidth]{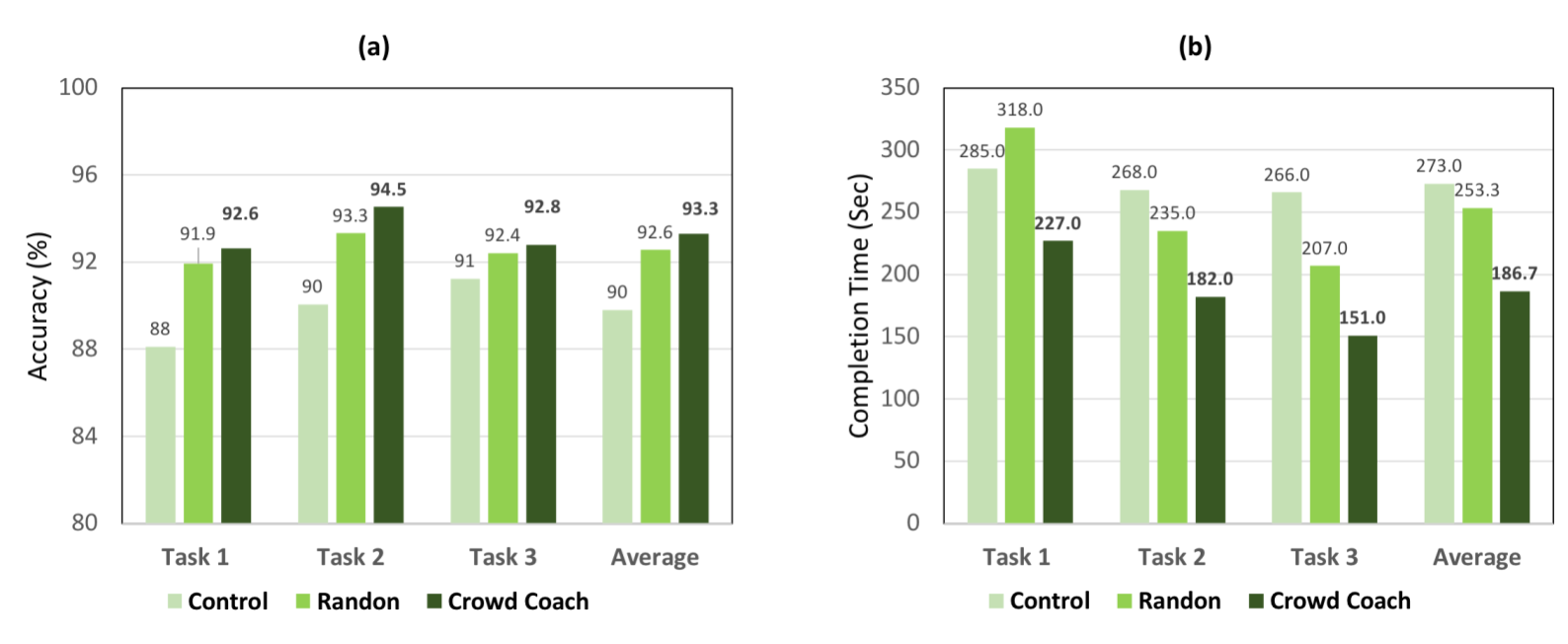}
  \caption{(a) Workers' accuracy for each audio transcription task when working with the different coaching methods.  (b) Workers' completion time for each task when working with the different coaching methods. \sys training outperformed other conditions in both accuracy and completion time.}~\label{fig:performance}
\end{figure}

During the study period, novice workers completed a total of 253 tasks across all 3 conditions. Some novice workers (13) decided to not complete all the tasks and dropped out. 26 workers remained in the control group finishing all three tasks, 26 workers in the random coaching group, and 25 workers in the \sys group. The retention rates per condition were similar (${\chi}^2$(2) = 0.03, p = 0.987). Figure \ref{fig:performance} presents the task accuracy (work quality) and the completion time of workers who completed all three tasks under different conditions. 

 We used an one-way MANOVA to compare whether workers' differences in task accuracy and completion time were significant across conditions. Notice that the one-way MANOVA helps us to determine whether there are any differences between independent groups (conditions) with more than one continuous dependent variables. In this case we have 3 independent groups (conditions) and 2 continuous dependent variables (task accuracy and completion time.)  The one-way MANOVA also helps us to study the two dependent variables at the same time. This is important as there might be a trade off between becoming faster at audio transcription tasks and workers' task accuracy. %completion time might be a trade-off to the task accuracy. Therefore, we have to consider the two variables at the same time.

The one-way MANOVA showed that there were significant differences in task accuracy and completion time across the 3 conditions (F(2,74) = 17.53, p < 0.0001). Given these significant differences that we discovered, we ran an univariate ANOVA per dependent variable (time and accuracy). We found that the difference were significant in completion time (F(2,74)= 54.18, p < 0.0001), but not significant in task accuracy (F(2,74)= 2.47, p = .09). Next, we conducted a Tukey test as post hoc analysis to understand more deeply the difference in time completion between conditions. We found that the competition time in the \sys condition (M = 184.1, SD = 12.36) was significantly less than the competition time in the control condition (M = 262.79, SD = 37.38) at p < 0.0001 and the competition time in the random coaching condition (M = 284.21, SD = 46.44) at p < 0.0001. 

Overall, our field experiment indicates that workers exposed to selected coaching snippets (\sys) were faster without sacrificing accuracy than workers without such coaching.

\subsection{Deployment Study}
We conducted a real world deployment of \sys to understand the type of work that is well or poorly supported by our system. We launched \sys and studied its use from June 25 to July 7th 2018. Similar to \cite{suzuki2016atelier,callison2014crowd}, we paid workers \$0.4 to install \sys in our deployment.

%\sys attracted interest from 179 workers, who successfully coached or received coaching for tens of tasks (see Table \ref{fig:deploymentTable}.) as a Chrome Extension 

\subsubsection{Deployment: Result}
\sys was installed by 179 workers, who successfully coached or received coaching for tens of tasks. 86\% of our participants were active users of the system (i.e., they either coached or micro-assessed the coaching); the rest used \sys more passively (they installed and potentially read the coaching.) 96 workers provided 363 coaching snippets, and 146 workers provided 1,401 micro-assessments to these coaching snippets. The median number of times workers coached each other was 2 and the worker who coached the most did it 32 times.

\begin{figure}
\centering
  \includegraphics[width=.6\columnwidth]{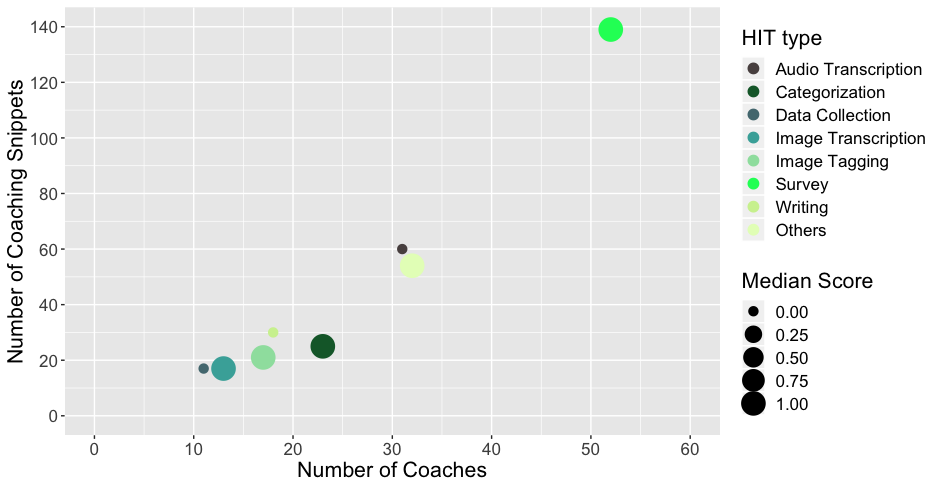}
  \caption{Overview of the total number of coaches, coach snippets generated for each HIT type and the median score the coaching received.}~\label{fig:deploy_overview}
\end{figure}

Figure \ref{fig:deploy_overview} and Table \ref{fig:deploymentTable} present an overview of how workers used \sys across different types of tasks. Each point represents a type of task for which workers used \sys. The X axis represents the total number of coaches who participated in a particular type of task. The Y axis represents the total number of coaching snippets generated for each task type. The size of each point is proportional to the average assessment score (upvotes minus downvotes) of the coaching snippets associated with a particular type of task. From Figure \ref{fig:deploy_overview} we observe that most coaches and coaching activity went into surveys and audio transcription tasks. Workers created a total of 139 coaching snippets for surveys, and 60 for audio transcription. %A majority of workers (26\%) decided to become coaches for survey tasks. 
We might be observing this preference for coaching surveys because they are one of the most common tasks on AMT and therefore workers are likely to have more experience with them. From Figure \ref{fig:deploy_overview} we also observe that all types of tasks had at least ten different coaches, and at least 20 associated coaching snippets that were in general rated positively by workers (the median score was above zero.)  

Workers overall tended to create different coaching snippets each time. Only 2\% of all coaching snippets were repeated verbatim. We observed that when workers shared the exact same coaching snippets, they usually tended to change the categorization of the coaching snippets. For instance, first they posted the coaching snippets in audio transcription tasks and then in surveys. In these cases, workers seemed to share ``general advice'' that they considered was relevant across tasks. An example of such coaching: \emph{``keep a list of requesters that you like completing hits for.''} In future iterations of \sys, we are considering enabling workers' to state when their coaching is general and applicable across AMT tasks. 

\begin{center}
\begin{table}
\begin{tabular}{ |c|c|c|c|} 
\hline
\tiny HIT Type & \tiny Coaching Snippets& \tiny Micro-Assessments&\tiny Micro-Assessments Score\\
\hline
%\multirow{3}{4em}{Multiple row} & cell2 & cell3 \\ 
\tiny Audio Transcription&
\tiny Total=60; Per worker: max =10; median =1 &
\tiny Total=176; Per worker: max =10; median =2&
\tiny Per task: min=-2; max =43; median =0\\ 

\tiny Categorization&
\tiny Total=25; Per worker: max =3; median =1&
\tiny Total=114; Per worker: max =8; median =2&
\tiny Per task: min=-1; max =25; median =1\\ 

\tiny Data Collection &
\tiny Total=17; Per worker: max =4; median =1&
\tiny Total=81; Per worker: max =4; median =3&
\tiny Per task: min=-2; max =22; median =0\\ 

\tiny Image Transcription &
\tiny Total=17; Per worker: max =3; median =1 &
\tiny Total=62; Per worker: max = 6; median =2&
\tiny Per task: min=0; max =16; median =1\\

\tiny Image Tagging  &
\tiny Total=21; Per worker: max =3; median =1&
\tiny Total=84; Per worker: max =7; median =3&
\tiny Per task: min=-1; max =16; median =1\\ 

\tiny Survey&
\tiny Total=139; Per worker: max =21; median =2 &
\tiny Total=406; Per worker: max =28; median =2&
\tiny Per task: min=-2; max =62; median =1\\ 

\tiny Writing &
\tiny Total=30; Per worker: max =3; median =1&
\tiny Total=114; Per worker: max =12; median =3&
\tiny Per task: min=-3; max =18; median =0\\ 

\tiny Others & 
\tiny Total=54; Per worker: max =6; median =1&
\tiny Total=364; Per worker: max =19; median =3&
\tiny Per task: min=-3; max =61; median =1\\ 
\hline
%\caption{Overview of the number of coaching snippets crowd workers generated for each type of task during our deployment.}
  %\vspace{-1.0pc}
%\label{fig:workflow}
\end{tabular}
\caption{Overview of the results from our deployment. Workers used \sys for a wide range of tasks. There was a tendency to use our system to coach especially survey and audio transcription tasks.}
\label{fig:deploymentTable}
\end{table}
\end{center}

From Table \ref{fig:deploymentTable} we observe that the coaching that received the highest maximum upvotes came from surveys, audio transcription, or tasks in the ``other'' category. Upon inspection, we identified that the coaching that workers upvoted the most tended to advise using smart interfaces to enhance one's work. For instance, one of the most upvoted coaching snippets involved advising workers to use an intelligent Chrome plugin: \emph{``Use a multi-highlite extension for chrome to hilite words that you need to make sure you see in a survey.''}.
Workers also tended to upvote coaching that shared best practices for working with specific requesters. The following was also one of the most upvoted coaching snippets: \emph{``With Stanford surveys remember to read carefully.''} %We believe that workers were likely upvoting more the coaching that provided more surprising 
The coaching that workers tended to downvote the most was coaching that appeared to have been categorized incorrectly. For instance, the following coaching was the most downvoted:\emph{``Some of these survey sites are legitimate but have either low pay for the work or only pay you in gift cards instead of cash.''} The coaching clearly seems to be for surveys tasks. Yet the coach said it was for audio transcription tasks. From Table \ref{fig:deploymentTable} we also observe that the coaching with most downvotes involved the ``other'' category which likely held a wide range of tasks. It is therefore probable that not all of the coaching in this category was relevant to workers. In the future we will explore allowing workers to correct the categories of coaching snippets. Coaching snippets that were also too specific tended to obtain the most downvotes. For instance, the following coaching snippet was one of the most downvoted and appears to be relevant just to the particular task the worker was doing:\emph{``with this count on no bonus. It is a game of chance''} %We also observed that workers assessed the quality of the coaching for each type of task differently. While the coaching given for the surveys was the one that received some of the most upvotes, the coaching for image transcription and image tagging was the best rated overall (the median score of these coaching snippets was XX.) Across tasks, all coaching generated in general positive assessments from workers (the median assessment score across tasks was positive.) %What we did observe was that workers tended to more rarely upvote or downvote a coaching snippet; but when they did, usually many others workers did the same. We believe this highlights that workers have the 

% \begin{figure}
% \centering
%   \includegraphics[width=.8\columnwidth]{usage.png}
%   \caption{Workers used Crowd Coach for a wide range of tasks. There was a tendency to use our system to coach especially survey and audio transcription tasks.}~\label{fig:deployment}
% \end{figure}

%\subsection{Post-Deployment Survey}
We believe that users' opinion and feedback are the foundation to design future iterations of \sys. We designed a post-deployment survey to collect opinions from real users, and posted the survey as a \$0.6 HIT on AMT available only to the workers who participated in our deployment. A total of 114 workers completed the survey, indicating a response rate of 64\%, and the median time to complete the survey was 8 minutes. Our survey revealed that workers overall found \sys useful (on a scale of 1-5, median =4, mode = 4). 42\% of the participants expressed that they felt they personally benefited greatly from having access to the coaching from others. In what follows, we present more details of the experiences with \sys that workers highlighted in the survey. We label worker participants as \#W.

{\bf Being Coached Facilitates Exploring New Work Areas.} Workers expressed that they used \sys to get a better sense of the experiences they would have if they did particular types of HITs. Being able to access such information seemed to help workers decide if they should venture into new work areas: \emph{``[What I liked the most about \sys is] receiving tips about tasks you hadn't performed yet, helping you decide if you should take them.''} (\#W21).

{\bf Coaching is a Social Good Activity to Help the Personal Growth of Peers.} Workers expressed that one of their main motivators for being a coach on our system was that they could help the personal growth of their peers: \emph{``Knowing that there's someone out there struggling to learn to ropes just like me and I can help them is what got me to keep providing tips when I can''} (\#W63). 

{\bf Coaching Brings the Best of External Sites into AMT.} Workers seemed to value that \sys allows them to stay on AMT while receiving important information they usually only obtained externally. \emph{``I think this is a very useful tool. In theory, it's faster and more useful than relying on an external site (like Turkopticon) to peruse the ratings of requesters.''} (\#W89). Most plugins for crowd work still require workers to spend significant time outside AMT. For example, workers can rate requesters on Turkopticon by giving scores from 1 to 5. But, if workers want to know more details about requesters, they have to go to an external website. Workers liked that this plugin allowed them to have conversation type interacts with their peers without leaving AMT: \emph{``I like the idea of interaction with fellow Turkers on the AMT site itself--- very similar to MTurk Suite, but with actual conversations. It lets you interact with people.''} (\#W102).

%\subsubsection{Tips are too general}
{\bf Struggle Between Too General or Too Specific Coaching.} Some workers felt that the system should guide coaching that was more specific rather than general: \emph{``Make sure people are commenting on a particular hit and not just offering general advice ... like ``getting a laptop'', or that `Wharton is usually a good one', I saw those on a lot of hits''} (\#W6). While other workers argued that the coaching should be more general so that novice workers could learn from them: \emph{``[...] I ended up leaving a couple of general tips--primarily useful for newbies.''} (\#W102).

\section{Discussion and Future Work}
We conducted a field experiment and a real world deployment to study different aspects of how \sys facilitated workers' personal skill development. Our field study experiment suggested that peer coaching can help novice crowd workers to improve in audio transcription tasks their speed while maintaining work quality. We believe that systems like \sys can likely empower crowd workers to increase their hourly pay as it could help workers to do more HITs per hour without sacrificing quality. 

It is expected that in the future, crowd markets will become employment hubs where increasingly large numbers of workers with varying expertise and skill levels compete for employment and contribute to projects \cite{kittur2013future}. It therefore becomes important to envision mechanisms that help the workers of these platforms to continuously grow to obtain better opportunities \cite{deng2013crowdsourcing}. 

Our field experiment suggested that we can use peer coaching to address the problem of facilitating personal skill development on crowd markets without requiring the insertion of experts. We also showed that for one of the most valued tasks on crowd markets--- audio transcription tasks--- even short coaching snippets that do not overwhelmingly distract from the task at hand can start to improve workers' speed. In the future, we would also like to explore using \sys to improve workers' personal skills for more complex professions such as CTOs, managers, or computer engineers. We also plan to explore the benefits of these type of crowd coach systems to facilitate skill development in rural communities where experts might be more scarce \cite{chiang2018exploring,angel2015participatory}.  

Research has demonstrated that peers can provide advice that is just as effective as advice from experts \cite{nelson2009nature,patchan2015understanding}. However, as with most pedagogy, the effects are not always consistent. Our field experiment and deployment can help researchers to better understand how peer advice plays out within the informal learning environment of crowd markets. Future work in this area could investigate more about how within crowd markets the coaching of peers differs from that of expert crowd workers %(e.g., super turkers \cite{}).% or under what circumstances is peer coaching most effective. 

Prior research has identified that the order of micro-tasks impacts performance \cite{cai2016chain}. For instance, having spaced repetitions usually impacts how many words a person can learn on their own for vocabulary acquisition \cite{edge2011micromandarin}; or mixing tasks with varied difficulty and similarity type impacts learning \cite{koedinger2012knowledge}. In the future, we plan to explore mixing peer coaching with the automatic generation of ``lists of tasks to do.''  Helping workers to select the tasks that might be most beneficial for them while pairing them with the best peer-coaching might enable workers to grow even more. 

\sys provides selected coaching snippets to novice workers for improvement. However, being exposed to peer coaching could also limit or shutdown workers' thought process and ideas on how to do a specific task. It is possible this could lead workers to believe that their approach is not best, as it is different than what others recommend, thereby discouraging workers from using creative approaches to solve tasks.  In the future, we are interested in exploring ways to enable coaching that helps workers to improve, but also promotes workers' initiative and creativity \cite{chan2016comparing}. We are  interested in exploring different reward schemes in this space. Similar to \cite{kulkarni2015peerstudio}, we are considering reward schemes used in domains like design where rapid iterations are prized \cite{buxton2010sketching,dow2009efficacy}. Our deployment also helped us to identify that \sys has the potential of assisting workers to venture into new work areas. Future work could explore how peer coaching might facilitate integrating minorities into areas where they have traditionally not been represented. 

In our real world deployment, we identified that workers tended to upvote coaching encouraging the use of intelligent interfaces to enhance their work. Crowd workers may not have been cognizant of existing technology that could support them in their work; and may have therefore appreciated learning about new tools to enhance their work. Alternately, it could also be that the people who are choosing to engage in crowd work have higher interests in new technologies, as previous research has also suggested \cite{brewer2016would, lampinen2016cscw}. Nonetheless, we believe there is value in exploring intelligent interfaces that support crowd workers in developing their personal skills. For example, while a substantial portion of online microtasks focus on creating training data sets for machine learning (to give just a few examples, \cite{braithwaite2016validating, miller2017parlai, harrison2016learning}),  in the future we plan to explore tools for helping workers to integrate machine learning into their workflow to have better work and learning experiences \cite{williams2016axis}, such as gaining a higher salary. 

Our real-world deployment also uncovered there were tensions between wanting too specific or too general coaching. Some workers preferred to have coaching that was tailored for the particular HIT they were doing; while other workers preferred more general advice. In the future we will explore interfaces that can facilitate labeling coaching into either ``specific'' or ``general'' to help crowd workers better control the type of coaching they receive. We noted that workers did not appear to value having ``mainstream'' or ``alternative'' type coaching. It seemed that knowing the generalizability of the coaching they received was more important. 
%Moreover, our real-world deployment result revealed that the tip which targets to a specific task is easily downvoted by the workers. However, in the post-deployment survey, the participants considered some of the tips cannot address well to the task content. Despite all of the displayed tips have high voting score, they are still not accurate enough to address all the same type HITs. It is possible that workers would upvote all the tips which they prefer. However, everyone's preferences are different. Therefore, the top voted tips are always the greatest common divisor of every workers' preferences. In other words, the tips which provide the general content easily got more upvote from the users. It is a dilemma for tip coaches to get more upvote or provide more accurate tips. Our future work is to find the how to get general but accurate tips.

Similar to other crowd powered systems \cite{huang2016there}, in the future, we also plan to run a longitudinal deployment of \sys to study the dynamics that emerge with workers as they use the system long term. We do not yet know whether workers may become reluctant to coach over time if they feel they are losing opportunities in so doing. For example, if a worker coaches about how to best deal with a particular requester, a greater supply of workers may then be able to successfully complete the work of such requester, thereby potentially crowding out the original Turker. Future work could thus explore the type of coaching that workers decide to limit (not share). There might also be opportunities to innovate and design new incentives to fill the void left by possible task or requester "hoarding" by leveraging the benefits of coaching. Future work could quantify the benefits of being a crowd coach long term. 

We also plan to study the system's sustainability. It is unclear whether workers will continuously have coaching to give each other, or if there is a finite number of tips that can be given. Crowd work is continuously evolving \cite{hara2018data}. We therefore believe that crowd workers will likely always have new advice to provide. We also believe that as workers evolve in their careers, some workers will be more likely to start assuming coaching roles and want to share their knowledge, regardless of what might exist previously. For example, Turkopticon was published in 2013 and it contains a lot of reviews of requesters \cite{irani2013turkopticon}. Still, crowd workers continually provide new reviews of requesters to it. In the future, we are also interested in exploring how peer advice could be used to help crowd workers directly make better wages, or learn how to delegate work for which they lack the skills to complete \cite{morris2017subcontracting}. 

We are also interested in the type of peer communication that workers will have with each other long term. Could there be a switch from communication focused on coaching each other to communicating focused on organizing labor unions? Prior work has shown how enabling communication between peers can lead to activism \cite{howard2016social}. It is unclear whether systems like \sys might facilitate workers' rights movements.

%As we saw in our survey, workers indicated that they found coaching their peers a helpful activity for improving themselves, such that coaching needn't be framed as a purely altruistic activity. 
%Future work could quantify the benefits of being a crowd coach, especially in the long term. 

{\bf Limitations.} The insights from this paper are limited by the methodology used and the population studied. While for our field experiment we used real tasks from crowd marketplaces and recruited real novice crowd workers, our skill development results might not yet generalize or apply to crowd work at large. Moreover, the pretest in our field experiment to guarantee that all the participants have similar skills means that we primarily evaluated the skill development benefits of our system with novices; we cannot speak to whether \sys could be suitable for more experienced workers, even though our ultimate goal is to provide a tool that can scale to different levels of experience. We did try however to address this issue by conducting a real world deployment where a range of hundreds of workers used our system on tens of different tasks. The deployment we conducted may also have novelty effects that need to be studied in greater depth in future work by conducting longitudinal studies. 

\section{Conclusion}
This paper introduced \sys, a Chrome extension to promote workers' personal skill building by supporting peer coaching while on the job. \sys uses peer micro-advice and a twofold reputation mechanism to help workers personally improve their skills. \sys differentiates itself from prior work by helping crowd workers to improve their skills by depending on the crowd workers themselves rather than requesters or experts. Moreover, \sys innovates on related work by embedding the help mechanisms within tasks so that work growth can be achieved while on the job, and by supporting workers to provide and receive the coaching through micro-coaching snippets rather than lengthy and comprehensive assistance. In this way, our system grounds itself in the practical concerns and constraints that both crowd workers and requesters face, such as limited time and resources available to devote to improving skills or task workflow designs. The present study sets the stage for future systems that focus on creating rewarding labor experiences where crowd workers can personally grow while on the job.

{\bf Acknowledgments.} This work was partially supported by FY2018 WV Research Challenge Grant award, J. Wayne and Kathy Richards Faculty Fellowship in Engineering and Research Grant from Leidos Laboratories. Special thanks AMT workers and the anonymous CSCW reviewers for their assistance in this research.

% Bibliography
\bibliographystyle{ACM-Reference-Format}
\bibliography{sample-bibliography}

\end{document}